%% file: main.tex
\documentclass[conference]{IEEEtran}
\IEEEoverridecommandlockouts
\usepackage[dvipsnames]{xcolor}
\usepackage{cite}
\usepackage{amsmath,amssymb,amsfonts}
\usepackage{algorithmic}
\usepackage{graphicx}
\usepackage{textcomp}
\def\BibTeX{{\rm B\kern-.05em{\sc i\kern-.025em b}\kern-.08em
    T\kern-.1667em\lower.7ex\hbox{E}\kern-.125emX}}

\usepackage{pgfplots}
\pgfplotsset{compat=1.18}
\usepackage[acronym]{glossaries}
\usepackage{adjustbox}
\usepackage{url}
\usetikzlibrary{positioning}
\usetikzlibrary{calc,backgrounds,patterns}
\pgfdeclarelayer{background}
\pgfdeclarelayer{foreground}
\pgfsetlayers{background,main,foreground}

\input{acronyms}

\newif\ifcomm

\ifcomm
\else
\commfalse
\fi
\ifcomm
\newcommand\jl[1]{\textcolor{violet}{JL: #1}}
\else
\newcommand\jl[1]{}
\fi

\begin{document}

\title{Direct Feature Access - Scaling Network Traffic Feature Collection to Terabit Speed
\thanks{This work was partly funded by the Bavarian Government through the Ministry of Science and Art through the HighTech Agenda (HTA).}
}

\author{\IEEEauthorblockN{1\textsuperscript{st} Lukas Froschauer}
\IEEEauthorblockA{\textit{Faculty of Applied Computer Science} \\
\textit{Deggendorf Institute of Technology}\\
Deggendorf, Germany \\
lukas.froschauer@th-deg.de}
\and
\IEEEauthorblockN{2\textsuperscript{nd} Jonatan Langlet}
\IEEEauthorblockA{\textit{EECS and Digital Futures} \\
\textit{KTH Royal Institute of Technology}\\
Stockholm, Sweden \\
jlanglet@kth.se}
\and
\IEEEauthorblockN{3\textsuperscript{rd} Andreas Kassler}
\IEEEauthorblockA{\textit{Faculty of Applied Computer Science} \\
\textit{Deggendorf Institute of Technology}\\
Deggendorf, Germany \\
andreas.kassler@th-deg.de}
}
\maketitle

\begin{abstract}
Real-time traffic monitoring is critical for network operators to ensure performance, security, and visibility—especially as encryption becomes the norm. AI and ML have emerged as powerful tools to create deeper insights from network traffic, but collecting the fine-grained features needed at terabit speeds remains a major bottleneck.
We introduce Direct Feature Access (DFA): a high-speed telemetry system that extracts flow features at line rate using P4-programmable data planes, and delivers them directly to GPUs via RDMA and GPUDirect —completely bypassing the ML server's CPU. DFA enables feature enrichment and immediate inference on GPUs, eliminating traditional control plane bottlenecks and dramatically reducing latency. We implement DFA on Intel Tofino switches and NVIDIA A100 GPUs, achieving extraction and delivery of over 31 million feature vectors per second—supporting 524,000 flows within sub-20~ms monitoring periods—on a single port. DFA unlocks scalable, real-time, ML-driven traffic analysis at terabit speeds, pushing the frontier of what is possible for next-generation network monitoring.

\end{abstract}


\input{introduction}

\input{related-work}
\input{design}

\input{implementation}
\input{evaluation}
\input{discussion}
\input{conclusion}

\vspace{-0.4em}
\bibliographystyle{IEEEtran}
\bibliography{resources}

\end{document}

%% file: acronyms.tex
\newacronym{ai}{AI}{artificial intelligence}
\newacronym{cpu}{CPU}{central processing unit}
\newacronym{dfa}{DFA}{Direct Feature Access}
\newacronym{dpi}{DPI}{Deep packet inspection}
\newacronym{dta}{DTA}{Direct Telemetry Access}
\newacronym{gpu}{GPU}{graphics compute unit}
\newacronym{gdr}{GDR}{GPUDirect RDMA}
\newacronym{int}{INT}{in-band network telemetry}
\newacronym{ml}{ML}{machine learning}
\newacronym{p4}{P4}{Programming Protocol-Independent Packet Processors}
\newacronym{qoe}{QoE}{quality of experience}
\newacronym{rdma}{RDMA}{remote direct memory access}

\newacronym{asic}{ASIC}{application specific integrated circuit}
\newacronym{ids}{IDS}{intrusion detection system}

\newacronym{sram}{SRAM}{static random access memory}
\newacronym{rocev2}{RoCEv2}{RDMA over converged ethernet version 2}
\newacronym{tam}{TAM}{traffic aggregation matrix}

\newacronym{iat}{IAT}{inter-arrival time}
\newacronym{lpm}{LPM}{longest prefix match}

\newacronym{dut}{DUT}{device under test}
\newacronym{snic}{SmartNIC}{smart network interface card}
\newacronym{ram}{RAM}{random sccess memory}
\newacronym{dpu}{DPU}{data processing unit}
\newacronym{alu}{ALU}{arithmetic logic unit}
\newacronym{pcie}{PCIe}{peripheral connection interface express}
\newacronym{numa}{NUMA}{non uniform memory architecture}
\newacronym{dma}{DMA}{direct memory access}

\newacronym{hta}{HTA}{High-Tech Agenda}

%% file: introduction.tex
\vspace{-0.4em}
\section{Introduction} \label{sec:introduction}
\vspace{-0.3em}
The ability to monitor network traffic in real time is critical for ensuring performance, security, and operational visibility. However, as global network traffic and data center traffic volumes continue to surge and encryption becomes the norm, traditional monitoring techniques are increasingly falling short. Operators face growing challenges in extracting meaningful insights from traffic at scale, particularly when fine-grained visibility is essential for applications like anomaly detection, flow classification, and \gls{qoe} estimation.

Traditional flow-level network telemetry collection approaches such as NetFlow~\cite{claiseCiscoSystemsNetFlow2004}, sFlow~\cite{wangSFlowResourceefficientAgile2004}, and IPFIX~\cite{aitkenSpecificationIPFlow2013}  rely on sampling and coarse feature sets to manage overhead, sacrificing granularity and limiting their effectiveness.
This limits the number of flows to monitor and their accuracy~\cite{seufertMarinaRealizingMLDriven2024}. 
Further, the set of monitored traffic features is limited by necessity, thus restricting the expressiveness of generated reports, significantly limiting the insights that operators gain. For example, without fine-grained, real-time insights into encrypted traffic patterns, a video streaming provider might fail to detect early indicators of \gls{qoe} degradation caused by transient congestion, leading to customer \mbox{dissatisfaction and churn.}

\gls{dpi}, while historically used to provide richer insights, has become less viable due to the ubiquity of encrypted traffic and the relentless increase in link speeds. Consequently, many operators now turn to \gls{ml}-based approaches, which can infer critical traffic properties using extracted flow-level features as they can provide greater flexibility and simplify deployment~\cite{grayHighPerformanceNetwork2021} while also augmenting classic solutions~\cite{salmanReviewMachineLearning2020}.

However, even ML-driven telemetry faces serious scalability challenges. 
Out-of-band feature extraction, which mirrors traffic to external servers, introduces significant reactivity latency and overhead. 
In-band feature extraction using programmable data planes (e.g., via P4~\cite{P4}) offers an alternative~\cite{grayHighPerformanceNetwork2021}, but existing solutions like Marina~\cite{seufertMarinaRealizingMLDriven2024} suffer from high control plane synchronization delays.
In-switch inference approaches~\cite{busse-grawitzPForestInNetworkInference2022,zhengIIsyPracticalInNetwork2022} are similarly limited by the restricted computational resources of programmable switches, allowing only simple ML models with limited accuracy and use cases.

On the other hand, operators increasingly aim to collect real-time and fine-grained network telemetry from the network fabric itself. For example, \gls{int} embeds telemetry data directly into packet headers~\cite{INT}, enabling real-time, hop-by-hop monitoring of network performance while providing detailed insights (e.g., latency, congestion, device behavior) without the extra overhead associated with traditional out-of-band monitoring. However, \gls{int} introduces challenges, including packet overhead and high processing latency of the telemetry data on standard \gls{cpu}~\cite{langletDirectTelemetryAccess2023}. Approaches such as \gls{dta}~\cite{langletDirectTelemetryAccess2023} ship network telemetry to external Collectors directly from the data plane using \gls{rdma}. This enables high-throughput, low-latency telemetry processing at scale due to reduced processing overhead, allowing for timely insights and adaptive responses to evolving traffic patterns. In contrast to Marina, \gls{dta} is limited to reduced telemetry items, does not extract and collect flow features (like packet sizes and inter-packet gaps) required for \gls{ai}-driven threat detection, flow classification, or application performance monitoring ~\cite{dimopoulosMeasuringVideoQoE2016,mazharRealtimeVideoQuality2018}, necessitating complementary approaches. 

In short, there is a growing need for telemetry systems that can: (i) extract rich, expressive flow features at line rate, (ii) deliver them with minimal latency to powerful ML engines, and (iii) operate at terabit-scale speeds \mbox{without \gls{cpu} bottlenecks.}

To address these limitations, we propose \gls{dfa}—a novel telemetry system that extracts flow features directly within P4-programmable switches and transfers them to ML inference servers' GPU memory using \gls{rdma} and \gls{gdr}~\cite{nvidiacorporation1OverviewGPUDirect}, completely bypassing the \gls{cpu}. \gls{dfa} seamlessly integrates the fine-grained feature extraction capabilities of Marina with the low-latency telemetry transfer of DTA~\cite{langletDirectTelemetryAccess2023}, enabling scalable, real-time network monitoring at unprecedented speeds.

\textbf{Our main contributions are:}
\begin{itemize}
    \item We design and implement \gls{dfa}, enabling direct, high-speed transfer of in-network extracted features to GPU memory for immediate ML inference.
    \item We integrate and extend Marina and \gls{dta} frameworks to achieve scalable telemetry with minimal overhead and no \gls{cpu} involvement.
    \item We evaluate \gls{dfa} on a real-world testbed comprising Intel Tofino switches and NVIDIA A100 GPUs, achieving extraction and delivery of over 31 million feature vectors per second, corresponding to 524,000 monitored flows under 20 ms intervals on a single network port.
\end{itemize}

\vspace{-0.15em}
The rest of the paper is structured as follows. In Section~\ref{sec:related-work}, we recap related work. Section~\ref{sec:design} presents \gls{dfa}'s system design. In Section~\ref{sec:implementation}, we outline the implementation of the three main \gls{dfa} components. Section~\ref{sec:evaluation} presents our testbed and evaluation results. Finally, Section~\ref{sec:conclusion} summarizes the paper and identifies future work. 

%% file: related-work.tex
\vspace{-0.45em}
\section{Background and Related Work} \label{sec:related-work}
\vspace{-0.5em}
Traditional flow-level monitoring solutions such as NetFlow~\cite{claiseCiscoSystemsNetFlow2004}, sFlow~\cite{wangSFlowResourceefficientAgile2004}, and IPFIX~\cite{aitkenSpecificationIPFlow2013} offer scalable but are limited in expressiveness and granularity, often relying on sampling and coarse export intervals. Typical use cases include estimation of flow sizes and their distribution or heavy hitter detection. More advanced analytics typically require \gls{dpi}, which copies user packets to inspection nodes. While \gls{dpi} provides fine-grained insights, it results in high overhead and is increasingly ineffective due to widespread traffic encryption.

To regain visibility, \gls{ml}-based monitoring techniques have emerged that require the extraction of feature vectors from user traffic and the analysis of such vectors by \gls{ml}-pipelines. In general, extraction and analysis can be done in-network or out-of-band on dedicated servers. When extraction is done in-network and analysis on external servers, feature vectors are synced to the control plane and sent to external servers for further analytics, leading to high overhead. Such approaches (i.e., Marina~\cite{seufertMarinaRealizingMLDriven2024}) leverage moment-based feature vectors, which enable the application of complex models on external \gls{ml} servers for high accuracy. However, they suffer from high overhead and latency (e.g., 500 ms) as the synchronization to the control plane and the transfer of the statistics to the \gls{ml} servers' \gls{cpu} is the bottleneck. In contrast, recent efforts in in-network \gls{ml}~\cite{busse-grawitzPForestInNetworkInference2022,xavierMAP4PragmaticFramework2022,zhengPlanterSeedingTrees2021} attempt to integrate analytics directly into programmable switches. While effective for lightweight tasks, such systems are constrained by the data plane's limited computational capabilities. In light of these limitations, Ilsy~\cite{zhengIIsyPracticalInNetwork2022} includes a second, larger \mbox{model on a dedicated server.}

Approaches like~\cite{lotfollahiDeepPacketNovel2018,draper-gilCharacterizationEncryptedVPN2016} demonstrate the feasibility of encrypted traffic classification on external servers, while others~\cite{dimopoulosMeasuringVideoQoE2016,mazharRealtimeVideoQuality2018,wassermannSeeWhatYou2019} focus on \gls{qoe} estimation and anomaly detection using statistical features. For example, Dimopoulos et al.~\cite{dimopoulosMeasuringVideoQoE2016}  and Mazhar and Shafiq~\cite{mazharRealtimeVideoQuality2018} present \gls{ml}-based frameworks for encrypted video \gls{qoe} inference, relying on traffic features extracted from network-level data. Wassermann et al.~\cite{wassermannSeeWhatYou2019} implement a low-latency video \gls{qoe} inference system based on traffic features. These systems, however, often depend on offline processing or simplified traffic representations, limiting their applicability \mbox{in high-throughput environments.}

\gls{int}~\cite{INT} is an advanced network monitoring technique that embeds telemetry data within the actual data plane inside packet headers, providing granular insights into network behavior. \gls{int} enables real-time collection of hop-by-hop network state information, including switch traversal latency, congestion information, and packet drop statistics. Unlike NetFlow or SNMP-based approaches, \gls{int} provides per-packet visibility into the network, offering finer granularity in performance monitoring. Capturing information related to nearly all flows is crucial to ensure fine-grained traffic analysis. However, programmable switching \gls{asic}s have severe resource constraints, which limit their ability to record all flows within a network. 

Several solutions for flow-level monitoring have been proposed that aim to collect similar statistics as NetFlow~\cite{10.5555/2930611.2930632, 10.1145/3190508.3190558, 9495221, 265057}. To mitigate memory requirements at the switch level, researchers used probabilistic data structures to store counters and track only heavy-flow IDs~\cite{10.1145/3230543.3230544}
, utilize signal-processing techniques to optimize resource usage~\cite{265057} or encode flow IDs and their corresponding counters directly within the \gls{asic}~\cite{10.5555/2930611.2930632}. Alternatively, \cite{10.14778/3467861.3467868,10.1145/3626775} are offloading flow IDs to the control plane while maintaining only counters in the data plane. \gls{p4} enables telemetry data to be pre-processed within the network itself, reducing the need for large-scale data transfers to external processing units~\cite{9064137}. Transmitting telemetry data to centralized collection systems can occur in-band ~\cite{Kim2015InbandNT, 265015,10.1145/3387514.3405877}, push-based~\cite{Cisco, Juniper, Telemetry,langletDirectTelemetryAccess2023}, poll-based~\cite{9447791, 10.1145/3387514.3406214}, or aggregated on-switch ~\cite{10.1145/3098822.3098829, 9064137}, each having different drawbacks on latency, scalability and overhead.

While prior work has advanced both telemetry collection and ML-driven network analytics, existing systems either suffer from high latency, limited computational power, or inadequate feature richness. Our work, Direct Feature Access (DFA), builds upon these foundations by combining Marina’s feature extraction capabilities with \gls{dta}’s low-latency telemetry export, enabling high-speed, fine-grained telemetry extraction directly into \gls{gpu} memory for immediate \gls{ml} processing—bridging the gap between in-network programmability and scalable, real-time \gls{ai} analytics.

%% file: design.tex
\section{System Design} \label{sec:design}
This section outlines the architecture and key components of our \gls{dfa} telemetry system, which is designed to support efficient, low-latency feature extraction from user traffic and export feature vectors to external \gls{ml} servers for \gls{ml}-based analysis. Our system builds on Marina's data plane feature extraction capabilities and utilizes extended  \gls{dta} capabilities to export features to an external ML server's \gls{gpu}s. The system consists of three components: \gls{dfa} Reporters, \gls{dfa} Translators, and \gls{dfa} Collectors. 

\begin{figure}
    \centering
    \includegraphics[width=\linewidth]{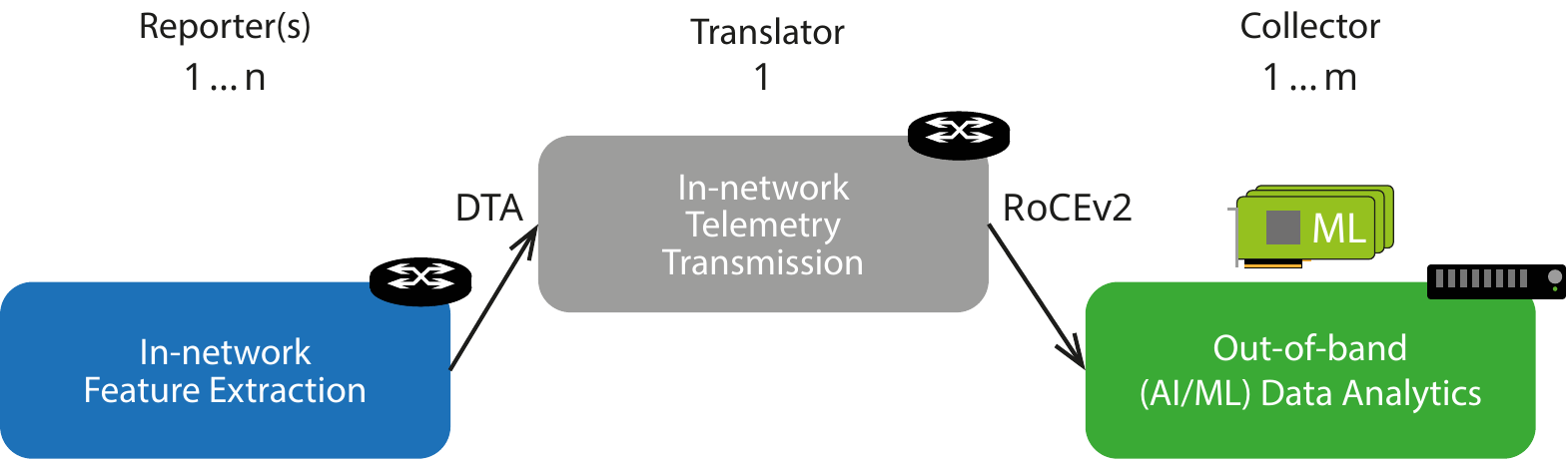}
    \vspace{-2.0em}
    \caption{An overview of the DFA telemetry data collection system, comprised of multiple instances of DFA Reporters, Translators, and Collectors.}
    \vspace{-1.5em}
    \label{fig:overview}
\end{figure}

In short, \gls{dfa} Reporters extract flow features from user traffic, enrich those features with moment-based statistics, and ship these enriched features to \gls{dfa} Translators using an extension of the \gls{dta} protocol. \gls{dfa} Translators receive \gls{dta} messages with features and create \gls{rocev2} messages to send these features to \gls{gpu}s, \gls{gpu} Memory on \gls{ml} servers. \gls{dfa} Collectors store a history of up to 10 feature vectors per flow, build derived features on CUDA cores, and trigger ML-inference tasks in \gls{gpu}s. Figure \ref{fig:overview} provides an overview of the \gls{dfa} system.

\vspace{-0.2em}
\subsection{DFA Reporter}
\vspace{-0.3em}
\gls{dfa} Reporters are \gls{p4} programmable switches, which forward the traffic on the network and extract basic flow-level information from user packets, including five-tuple, packet size, and \gls{iat}.  In addition to logging those basic parameters, \gls{dfa} Reporters perform lightweight computations on them to build feature vectors commonly used in \gls{ml} inference. These include the sum of \gls{iat}, its squared and cubed sum, sum of packet sizes, and its squared and cubed form. Per flow feature vectors are temporarily stored in stateful \gls{sram} registers on the \gls{dfa}~Reporters data plane.

The \gls{dfa} Reporter combines the functionalities of the Marina data plane and the \gls{dta} Reporter. 
The primary motivation for merging both systems is to bypass Marina's data export mechanism, which relies on periodic \gls{dma} operations to transfer all data from the data plane to the control plane before exporting it to an external \gls{ml} server. The \gls{dma} process alone takes circa 268~ms, whereas \gls{dta}'s mechanism is constrained only by the network delay~\cite{seufertMarinaRealizingMLDriven2024}.

The \gls{dfa} Reporter is designed to support per-flow configurable telemetry reporting intervals directly from the data plane to a \gls{dfa} Translator for further processing, thus avoiding costly sync to the control plane. In more detail, each flow is associated with a tracking table that records the time when the last feature vector of that given flow has been sent to the Translator. Suppose the current time and the last reported interval are larger than the flow monitoring period. In that case, telemetry reporting is triggered, creating a feature vector for the given flow sent to the \gls{dfa} Translator. When a send operation is triggered, the current traffic packet is cloned within the switch. Its original content is then removed and replaced with feature data to create the report. Moreover, the \gls{dfa} design enables multiple Reporters to be deployed within a single network, all capable of exporting data \mbox{to one or more \gls{ml} servers.}

While the \gls{dfa} Reporter supports fine-grained telemetry reporting intervals directly from the data plane that are customizable \footnote{\gls{dta} uses a probabilistic method to initiate data export.},  Marina has a global monitoring interval of around 500~ms and requires costly \gls{dma} copy operations to sync the registers collecting the telemetry data from the user packets to the control plane. In addition, Marina requires another protocol (e.g., TCP) to send feature vectors from the control plane to the \gls{cpu} of \gls{ml} servers, further contributing to long monitoring intervals.  While such large monitoring intervals may be acceptable for some use cases, such as \gls{qoe} estimation, our design aims to support much smaller monitoring intervals of up to 20~ms for immediate ML-inference triggering. Such low-latency \gls{ml} inference may enable new use cases not supported by the state-of-the-art. For example,\cite{287396} builds a robust traffic representation in the form of a fine-grained traffic aggregation matrix, which can abstract critical flow features, at around 40~ms intervals, well supported by our design. 

\vspace{-0.2em}
\subsection{DFA Translator}
\vspace{-0.3em}

The \gls{dfa} Translator serves as a central aggregator of telemetry reports to reduce the processing burden on the \gls{ml} server. It abstracts the \gls{rdma} process of the \gls{dfa} Collector, managing the connection and orchestrating efficient data transfers. The final hop in the data path\textemdash between the \gls{dfa} Translator and the \gls{dfa} Collector\textemdash leverages \gls{gdr}. As previously outlined, \gls{rdma} bypasses host compute resources and significantly lowers latency, particularly for small payload sizes. 

Using a round-robin mechanism, the \gls{dfa} Translator selects a suitable memory address in the address space provided by the \gls{dfa} Collector. The selected address depends on the flow ID and the history entry associated with the current report. Which history entry the report carries is determined by an 8-bit register, with one register entry per flow ID. This register acts as a counter that is readable and writable from the data plane, resetting to 0 when the maximum history index is reached.

A lightweight transport protocol is employed between the \gls{dfa} Reporter and the \gls{dfa} Translator to minimize the network overhead introduced by the telemetry system and enhance scalability. Specifically, the data transmission is based on a derivative of the \gls{dta} Key-Write primitive, which is optimized for key-value pair operations and supports efficient, low-overhead message transport~\cite{langletDirectTelemetryAccess2023}. This mechanism allows for simple and fast encoding of telemetry data (i.e., Marina feature vectors)  suitable for high-rate export \mbox{from programmable devices.}

\begin{figure}
    \centering
    \begin{adjustbox}{width=\linewidth}
	\includegraphics[width=\linewidth]{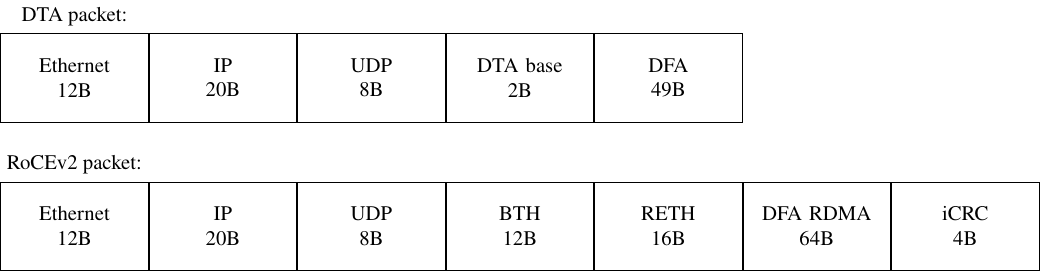}
    \end{adjustbox}
    \vspace{-2.0em}
    \caption{Packet headers and sizes for DTA (used by the DFA Reporter) and RoCEv2 packets with DFA data (used by the DFA Translator).\jl{Larger font?}}
    \vspace{-1.5em}
    \label{fig:dta-rocev2-comparison}
\end{figure}

Figure~\ref{fig:dta-rocev2-comparison} illustrates the headers and their sizes for a \gls{dta} packet when used with our \gls{dfa} data (created by the \gls{dfa} Reporter and parsed by the \gls{dfa} Translator) and an \gls{rocev2} packet with \gls{dfa} as payload (created by the \gls{dfa} Translator and parsed by the \gls{dfa} Collector). The \gls{dta} base header contains the flow ID and flags describing the following data header. The sizes for the \gls{dfa} data header and \gls{dfa} \gls{rdma} payload are different because \gls{rocev2} only supports payloads fitting in factors of two (e.g., 64). Thus, the \gls{dfa} \gls{rdma} header contains extra padding \mbox{to extend the \gls{rdma} payload to 64 Byte.}

\vspace{-0.2em}
\subsection{DFA Collector}
\begin{figure}
\centering
\includegraphics[width=.9\linewidth]{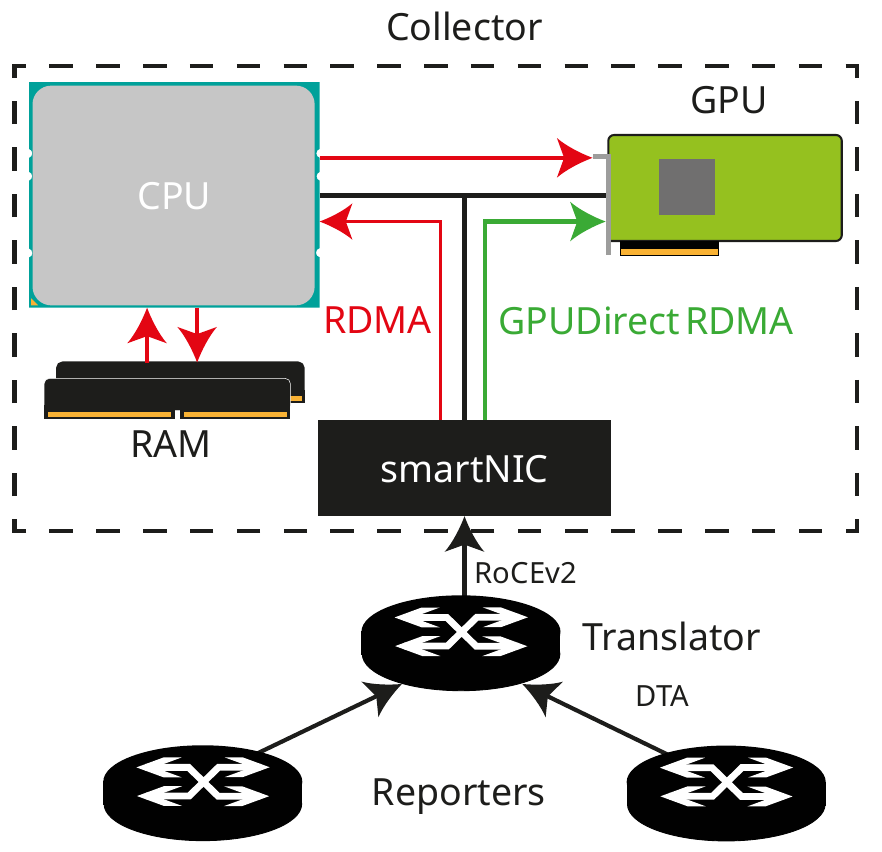}
\vspace{-1.0em}
\caption{Schematic path of data through Collector. DTA (in red) copies data from the smartNIC to host memory and requires costly memcopy operations to the GPU involving the host CPU for further processing. In contrast, the DFA Collector (in green) uses GPUDirect RDMA to bypass the host CPU.}
\vspace{-0.8em}
\label{fig:collector-data-path}
\end{figure}

\vspace{-0.3em}
The \gls{dfa} Collector is the system's telemetry data sink. It exposes a memory region accessible via \gls{rdma}, which can reside in either system memory or directly in \gls{gpu} memory depending on the deployment configuration. This flexibility allows for efficient data ingestion tailored to the processing backend. By bypassing the \gls{cpu} and enabling direct memory access, the \gls{dfa} Collector minimizes latency and maximizes throughput. The differences for the data path with classic \gls{rdma} into system memory and a copy operation to the \gls{gpu}, and \gls{gdr} are shown in Figure~\ref{fig:collector-data-path}. To facilitate the \gls{rdma} connection, the Collector also shares its \gls{rdma} connection details with the \gls{dfa} Translator. This is done through the \gls{rdma} connection manager, included in Infiniband libraries. The connection manager provides access to \gls{rdma} queue-pairs, which facilitate access rights to external devices. Depending on the access rights granted, external devices can read from or write to the memory address space provided. Our system provides write access to the Translator \mbox{through a single \gls{rdma} queue-pair.}

\begin{figure}
    \centering
\begin{adjustbox}{max width=\linewidth, max height=5cm,keepaspectratio}
\includegraphics{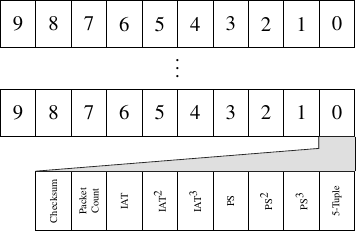}
\end{adjustbox}
    \vspace{-0.8em}
    \caption{Illustration of the memory structure maintaining 10 history entries per flow record. Each entry comprises telemetry features (packet count, inter-arrival time (IAT), packet sizes (PS)), the network five-tuple identifier, and a checksum for flow identification.}
    \vspace{-1.4em}
    \label{fig:memory-structure}
\end{figure}

In Figure~\ref{fig:memory-structure}, the memory structure employed in the Collector is shown. In addition to the telemetry data, each entry contains the flow ID and the five-tuple associated with the entry. This is done to keep all data in one place and allow the system to identify the data and flow without additional synchronization. 

\begin{table*}
    
    \vspace{-0.8em}
    \caption{Statistics and Meta-Information gathered at line-rate and transmitted through RDMA~\cite{seufertMarinaRealizingMLDriven2024}. $log^*$ denotes the approximation of the logarithm based on lookup tables.}
    \label{tab:telemetry-statistics}
    \centering
    \begin{tabular}{|c|c|c|c|}
    \hline
    \textbf{Name}   & \textbf{Implementation}   & \textbf{Size}   & \textbf{Description} \\
    \hline
    Packet Count    &  & 4B    & Basic statistic to calculate ratio per packet. \\
    Last packet timestamp   &   & 4B    & Determine when the last packet has arrived and calculate IAT internally. \\
    \hline
    Sum of Inter-Arrival Time (IAT)    & $\sum log^*$ (IAT) & 4B    & Calculate mean of IAT. \\
    Sum of IAT\textsuperscript{2}  & $\sum log^*$ (IAT\textsuperscript{2})  & 4B    & Variance, standard derivation, coefficient of variation \\
    Sum of IAT\textsuperscript{3}  & $\sum log^*$ (IAT\textsuperscript{3})  & 4B    & Skewness \\
    \hline
    Sum of Packet Size (PS)    & $\sum log^*$ (PS)  & 4B    & Volume, volume ratio, mean of PS \\
    Sum of PS\textsuperscript{2}   & $\sum log^*$ (PS\textsuperscript{2})   & 4B    & Variance, standard derivation, coefficient of variation \\
    Sum of PS\textsuperscript{3}   & $\sum log^*$ (PS\textsuperscript{3})   & 4B    & Skewness \\
    \hline
    Five-Tuple  &   & 17B   & src + dst IP addresses, src + dst layer 4 ports, layer 4 protocol \\
    \hline
    \end{tabular}
    \vspace{-2.0em}
\end{table*}

%% file: implementation.tex
\section{Implementation} \label{sec:implementation}
\vspace{-0.3em}
Both the Reporter and Translator are divided into control plane and data plane components, with the control plane implemented in Python and the data plane written in \gls{p4}. The Collector, responsible for receiving and processing telemetry data, is developed using C++ and CUDA.

\vspace{-0.2em}
\subsection{DFA Reporter}
\begin{figure}
    \centering
    \includegraphics[width=\linewidth]{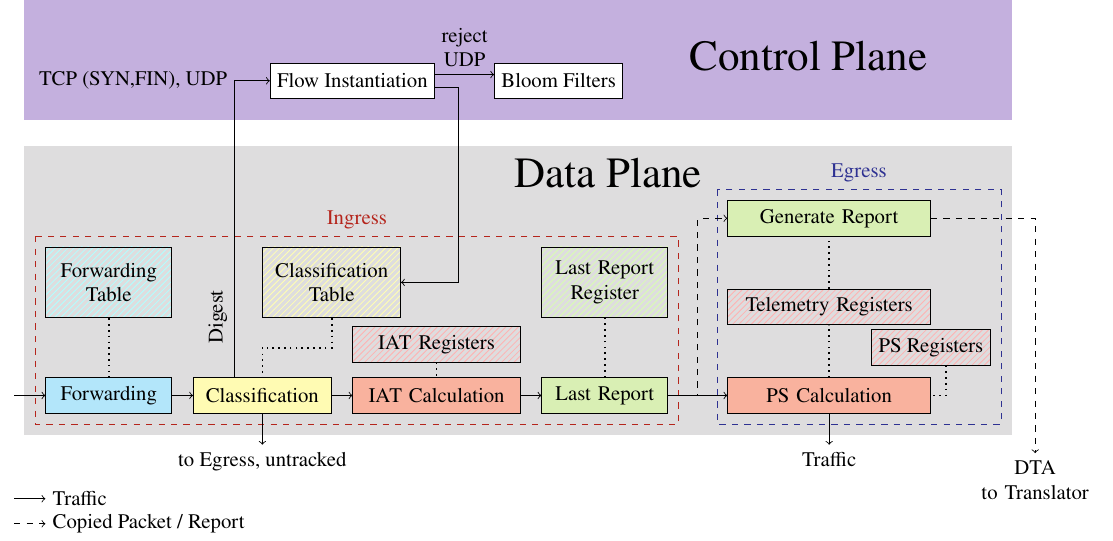}
    \vspace{-2.0em}
    \caption{DFA Reporter design overview indicating interaction between data plane (implemented in P4) and control plane programs (implemented in Python).} 
    \vspace{-1.4em}
    \label{fig:reporter-overview}
\end{figure}

\vspace{-0.3em}
The \gls{dfa} Reporter combines the \gls{dta} Reporter and the Marina data plane, which was converted from \gls{p4}\textsubscript{14} to \gls{p4}\textsubscript{16} and integrated into the \gls{dta} telemetry pipeline. We visualize our reporting pipeline in Figure~\ref{fig:reporter-overview}. 
The \gls{dfa} Reporter is designed as an L3 forwarding device. Incoming packets are first forwarded when entering the switch. The control plane decides whether a flow is tracked by installing a tracking rule in the classification table, and thus, if features are extracted and stored in the switch's tracking registers. The classification table uses the five-tuple as its exact-match key and outputs a unique flow ID. The classification table has the capacity for 2\textsuperscript{17} (131,072) entries, the maximum supported flows in a single pipeline. 
A table miss indicates a new flow, triggering a digest message to the control plane, to let it decide whether to track the flow or not. To prevent congesting the interface to the control plane,  partitioned bloom filters are used in the data plane to filter out undesired UDP flows. These bloom filters are set through a separate counting bloom filter in the control plane, and they are used to prevent subsequent packets of a UDP flow from creating control plane notifications, as UDP packets don't signal the beginning and end of the flow. 
For TCP, only packets using SYN and FIN flags are forwarded to the control plane. Our work has no explicit mechanism for malicious or erroneous flows, like \mbox{a TCP flow ending without the FIN flag.}

Important moment-based statistics require calculating and tracking the \gls{iat} 
within a flow, along with packet sizes and their respective square and cube functions in the data plane. Due to limitations of the \gls{p4} language and the 32-bit register constraints of the hardware~\cite{seufertMarinaRealizingMLDriven2024}, the values are approximated through pre-populated match-action tables using logarithmic values, resulting in approximations of the actual values. 

The calculation of the telemetry features and the telemetry report generation are split across the ingress and egress of the switch. This separation is necessary due to resource constraints (e.g., \gls{sram}) and the availability of certain functions (e.g., packet cloning, timestamps) in only one section. To generate feature vectors, we first check whether sufficient time has passed since the last feature vector for a given flow. If so, we clone the original packet in the ingress traffic manager. We use a register array to track the timestamp when the last feature vector report was sent per flow. The cloned packet is then truncated, the telemetry data is read from the correct registers to create the feature vector in the egress. The feature vector includes the statistics listed in Table~\ref{tab:telemetry-statistics} and the flow ID. All telemetry data requiring packet timestamps (e.g., \gls{iat}) is captured for the time the packet enters the switch. Since ingress is the only place where entry timestamps are available, metadata bridging is used to transfer this data to the egress.

\vspace{-0.2em}
\subsection{Translator}
\vspace{-0.3em}
The \gls{dfa} Translator is a modification of the \gls{dta} Translator and similar to the Reporter, extends a standard L3 forwarding component. In contrast to \gls{dta}, however, 
the \gls{dfa} Translator supports the custom \gls{dfa} primitive based upon the key-write primitive and a mechanism for selecting multiple memory addresses for each history entry. The key-write primitive's single data field is replaced with the seven data fields of \gls{dfa} and the five-tuple for flow identification. 

The Translator consists of a Python control plane program that initiates the \gls{rdma} connection with the Collector and sets the forwarding rules. This control plane is less involved compared to the Reporter's. The data plane is implemented in \gls{p4}. It determines the memory address for each report on the Collector and supports \gls{rdma} congestion handling. Additionally, the Translator replaces the \gls{dta} headers with \gls{rocev2} headers to perform \gls{rdma} WRITE Only operations. 

\vspace{-0.2em}
\subsection{Collector}
\vspace{-0.3em}
The \gls{dfa} Collector extends the \gls{dta} Collector with a custom library to enable CUDA and \gls{gdr} functionality, allowing for minimal change to the overall program and structure. The Collector in this project is a simple implementation suitable for performing the tests shown in Section~\ref{sec:evaluation}. The custom library allows a drop-in replacement for the allocation and access to memory cells, as well as adding a custom CUDA kernel program for the bandwidth tests. Despite unified memory addressing through the \textit{nvidia-peermem} Linux kernel module, allocation and access of \gls{gpu} differ from the same operations on system memory. 

%% file: evaluation.tex
\vspace{-1.0em}
\section{Evaluation} \label{sec:evaluation}
\vspace{-0.3em}
To evaluate the performance and scalability of our system, we address the following key questions:
\begin{itemize}
    \item \textbf{Data Plane Resource Usage (Section \ref{eval:resource}):} How does integrating Marina’s feature extraction and storage into the data plane affect resource utilization on the switch?
    \item \textbf{Scalability (Section \ref{eval:feature-size}):} What is the impact of feature vector size on message rate and telemetry \mbox{throughput?}
    \item \textbf{Comparison of GPUDirect RDMA and RDMA with Memcopy (Section \ref{eval:rdma-comparison}):} Does \gls{gdr} improve data throughput over classic \gls{rdma}?
\end{itemize}

\vspace{-0.2em}
\subsection{Testbed Setup}
\vspace{-0.3em}
To answer the above questions, we set up a testbed in our lab that consists of two x86-based servers connected via QFSP28 fiber-optic transceivers and cables and linked through an Edge-Core DCS801 Wedge100BF-32QS \cite{edgecorenetworkscorporationDCS801Wedge100BF32QSDS2023} switch, creating 100Gbps links. The first server acts as a traffic generator, running T-Rex~\cite{TRex} on a dual Intel Xeon Gold 6326 @2.90 GHz system with 256 GB of \gls{ram}, Ubuntu 22.04, and a NVIDIA/Mellanox ConnectX-6 DX \gls{snic}. The second server acts as the \gls{dut} and is composed of a Dell EMC PowerEdge R7525 equipped with dual AMD EPYC 75F3 @2.95 GHz \gls{cpu}s, 512 GB of \gls{ram}, two NVIDIA A100 80 GB \gls{gpu}s, and runs Ubuntu 22.04. It includes a NVIDIA/Mellanox BlueField-2 \gls{dpu} (ConnectX-6 DX) operating in \gls{snic} mode.

\vspace{-0.2em}
\subsection{Data Plane Resource Usage} \label{eval:resource}
\vspace{-0.3em}
\gls{dfa} Reporter needs to store and track hundreds of thousands of flow features in the data plane on a \gls{p4} programmable target. This is challenging on networking hardware that operates at 6.4 Tbps~\cite{edgecorenetworkscorporationDCS801Wedge100BF32QSDS2023} due to the limited amount of resources available. While data plane devices such as Intel Tofino offer various hardware components for performing calculations and storing state within the data plane, the additional overhead of extracting and storing feature vectors may be significant. In order to answer the question of how the feature extraction and storage affect resource utilization of the switch, we focus our evaluation on the \gls{sram} and Stateful \gls{alu} usage, as these provide the essential building blocks for implementing data registers in the data plane that are used by \gls{dta} to store flow feature vectors. 

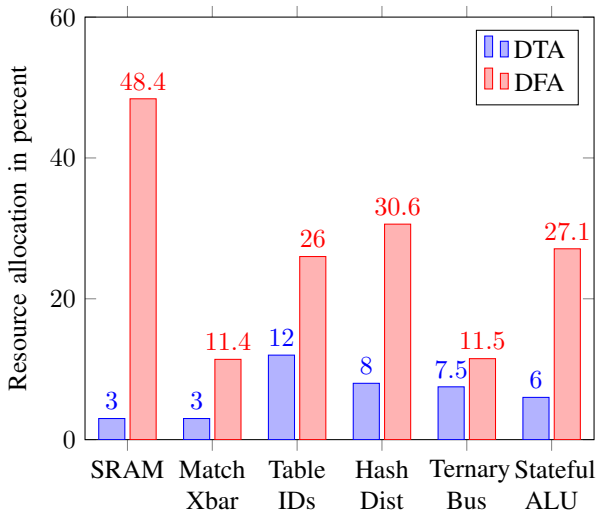
\begin{figure}
    \centering
    \begin{adjustbox}{width=0.9\linewidth}
    \begin{tikzpicture}
        \begin{axis} [ybar,
        symbolic x coords={SRAM, Match Xbar, Table IDs, Hash Dist, Ternary Bus, Stateful ALU},
        xtick=data,
        ymin=0,
        ymax=60,
        bar width=10pt,
        xticklabel style={
            text width=1cm,
            align=center
        },
        xticklabels={
            SRAM,
            Match\\Xbar,
            Table\\IDs,
            Hash\\Dist,
            Ternary\\Bus,
            Stateful\\ALU
        },
        ylabel={Resource allocation in percent},
        enlarge x limits=0.1,
        legend pos=north east,
        nodes near coords]

        \addplot coordinates {(SRAM,3) (Match Xbar,3) (Table IDs,12) (Hash Dist,8) (Ternary Bus,7.5) (Stateful ALU,6)};
        \addplot coordinates {(SRAM,48.4)  (Match Xbar,11.4)    (Table IDs,26)  (Hash Dist,30.6)    (Ternary Bus,11.5)    (Stateful ALU,27.1)};

        \legend{DTA, DFA}
            
        \end{axis}
    \end{tikzpicture}
    \end{adjustbox}
    \vspace{-1.0em}
    \caption{Comparison of resource footprint between \gls{dta}~\cite{langletDirectTelemetryAccess2023} and \gls{dfa}.}
    \vspace{-1.4em}
    \label{fig:resource-footprint}
\end{figure}

\begin{figure}
    \centering
    \includegraphics[width=\linewidth]{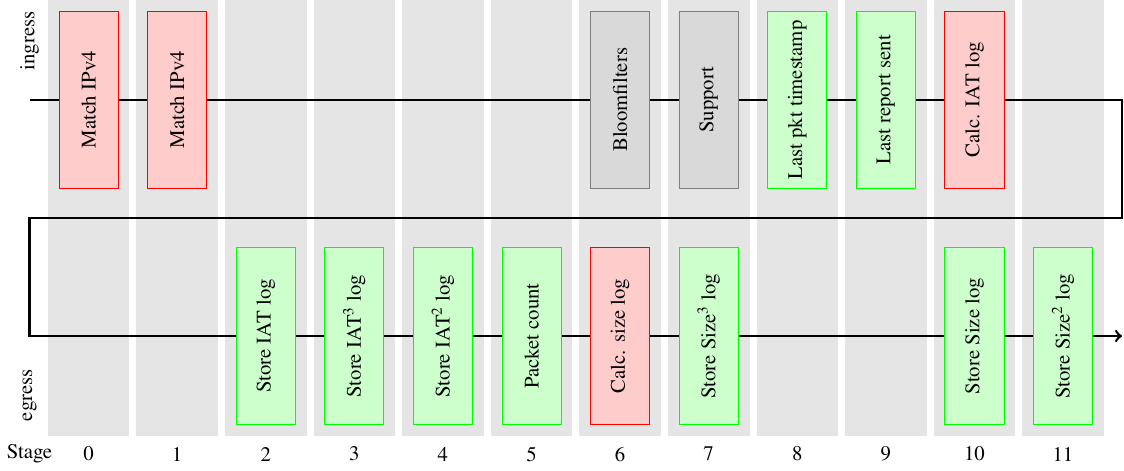}
    \vspace{-1.4em}
    \caption{Overview of registers, tables, and support functions placement in Intel Tofino stages.}
    \vspace{-1.5em}
    \label{fig:stage-overview}
\end{figure}

We use Intel’s \gls{p4} Insight tool~\cite{IntelP4Insight} and plot data plane resource allocation (percentage) in Figure~\ref{fig:resource-footprint}, comparing average resource usage of the Reporter program with \gls{dta} and \gls{dfa}. For \gls{dta}, we use data from~\cite{langletDirectTelemetryAccess2023}. Incorporating Marina's feature extraction and temporal storage into \gls{dta}'s Reporter substantially increases \gls{sram} usage, limiting availability of exact-match tables and particularly registers for other programs alongside \gls{dta}. Resource utilization also varies across pipeline stages, with the highest usage in the egress pipeline, where extracted features are stored in register arrays. Register and match-action table placements are shown in Figure~\ref{fig:stage-overview}. Of twelve total stages available on the \gls{p4} target, nine are filled with registers containing 2\textsuperscript{17} (131,072) 32-bit entries each. This also represents the number of flows trackable per switch pipeline. Our switch has two pipelines connected to physical ports, supporting 262,144 flows in total, or 131,072 per pipeline. Some Intel Tofino models with four pipelines support 524,288 flows. The maximum number of supported 32-bit entries in the Intel Tofino architecture is 143,360~\cite{seufertMarinaRealizingMLDriven2024}. 

\vspace{-0.6em}
\subsection{Data Plane Performance and Scalability} 
\vspace{-0.4em}\label{eval:feature-size}
In order to evaluate the scalability of our approach, we evaluate how many feature vectors \gls{dfa} can be sent from the data plane to the \gls{ml} servers \gls{gpu} using a single port. A key performance metric is the achievable \gls{rdma} data throughput between the data plane, where features are extracted temporarily, and the \gls{ml} servers \gls{gpu}, as it determines how many feature vectors can be transmitted to the \gls{ml} model per second. Potential bottlenecks in this connection include the 100 Gbps fiber-optic link and the \gls{pcie} interfaces of both the \gls{snic} and the \gls{gpu}s. 

To evaluate the scalability of our approach, we conducted several tests increasing the size of the feature vectors. We used the T-Rex traffic generator as a Reporter. The tests were conducted at fixed packet rates for each run, with different payload sizes (feature vector sizes) used across separate runs. During each 10-second test interval, the packet rate remained constant. A large contiguous memory buffer was pre-allocated on the receiving side of the Collector to accommodate all incoming packets without overwriting. The traffic generator incrementally updated the telemetry identifier for each packet, allowing the Translator to store each payload in a distinct memory cell. At the end of each run, the number of written and empty memory cells was recorded and compared to the number of telemetry packets sent. The check for the content of the memory cells is implemented in a CUDA kernel, allowing for millions of threads to check through all memory cells. This process was repeated five times per configuration. The message rate was considered validated if the mean discrepancy between the number of written memory cells and transmitted packets was less than 0.1\%. In total, a memory space for over 1 billion entries was kept available, in the \gls{gpu}memory, for tests up to 64 Byte payloads. For 128 Byte payloads, the maximum number was reduced to 500 million to stay within the memory size of the \gls{gpu}. On the 10-second test interval, this would have allowed for measurable message rates of up to 100~Mpps or 50~Mpps, depending on the tested payload size. We report the average number of messages received per second and the resulting bandwidth (in Gbps).

\begin{figure}
    \centering
    \begin{tikzpicture}
        \begin{axis}[ybar,
        symbolic x coords={8B, 16B, 32B, 64B, 128B},
        xtick=data,
        ymin=20,
        ymax=37,
        axis y line*=left,
        ylabel={Million messages per second},
        bar width=10pt,
        enlarge x limits=0.2,
        nodes near coords,
        nodes near coords align={vertical},
        width=0.8\linewidth]
            \addplot[xshift=-7pt,color=blue,fill=blue!30] coordinates {(8B,32) (16B,34) (32B,34) (64B,31) (128B,28)};
        \end{axis}

        \begin{axis}[ybar,
        symbolic x coords={8B, 16B, 32B, 64B, 128B},
        xtick=data,
        ymin=0,
        ymax=32,
        axis y line*=right,
        ylabel style={at={(axis description cs:1.1,0.5)}},
        ylabel={Payload Bandwidth in Gbps},
        bar width=10pt,
        enlarge x limits=0.2,
        nodes near coords,
        xlabel={RDMA payload sizes},
        width=0.8\linewidth]

        \addplot [xshift=7pt,color=red,fill=red!30] coordinates {(8B,2) (16B,4.4) (32B,8.7) (64B,15.9) (128B,28.7)};
            
        \end{axis}
    \end{tikzpicture}
    \vspace{-1.2em}
    \caption{Comparison of achievable message rates (in blue) and payload bandwidths (in red) using GDR at different payload sizes. A complete feature vector fits in a 64 Byte payload; however, a reduced or multiple batched could improve feature vector throughput.}
    \vspace{-1.8em}
    \label{fig:payload-size}
\end{figure}
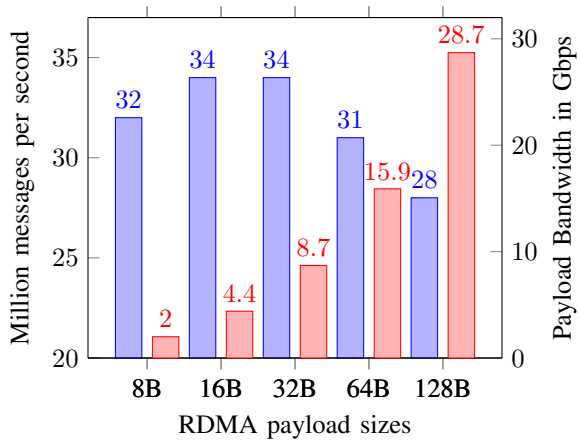

As can be seen in  Figure~\ref{fig:payload-size}, with an 8B payload size, \gls{dfa} can ship 32 million messages per second from the data plane directly to the \gls{gpu} memory of the Collector. However, Marina's full feature vector requires 45B. Consequently, transmitting full Marina feature vectors, \gls{dfa} can achieve around 31 million messages per second with a 64B \gls{rdma} payload, resulting in around 15.9 Gbps throughput. As shown in Figure~\ref{fig:payload-size}, message rates decrease with larger payload sizes. For example, at 128 Bytes, the achievable rate is around 28 million messages per second.

\vspace{-0.3em}
\subsection{GPUDirect/RDMA Comparison} \label{eval:rdma-comparison}
\vspace{-0.2em}
\gls{dfa} sends feature vectors directly to the \gls{gpu} memory bypassing host \gls{cpu} completely. This is different from both Marina and \gls{dta}. Marina sends feature vectors from the control plane to external ML servers \gls{cpu}, uses memory copy to the \gls{gpu} to trigger \gls{ml}-inference. \gls{dta}, on the other hand, sends telemetry directly to the host memory. However, further processing telemetry data with \gls{dta} using an ML-model would require copying telemetry data to the \gls{gpu} using costly GPU memcopy operations. 

In order to quantify the benefit of \gls{dta}'s approach, which sends feature vectors directly to the \gls{gpu}'s memory using \gls{gdr}, we compare \gls{gdr} with a baseline that receives 64 Byte feature vectors into the host memory using \gls{rdma} followed by a batched CUDA host-to-device memcopy operation, that copies the memory to the \gls{gpu}. While this test is similar to results from  Section~\ref{eval:feature-size}, we fix the payload to a single Marina feature vector (64 Byte payload size). We compare to  \gls{dta} performance using numbers found in ~\cite{langletDirectTelemetryAccess2023}. Note that~\cite{langletDirectTelemetryAccess2023} assumes an 8 Byte payload, and the system used for \gls{dta} tests differs from our system because our ML server is based on the AMD EPYC architecture. More specifically, the \gls{snic}, system memory, and the \gls{gpu} are connected to the same \gls{cpu}, but they are on different \gls{numa} nodes, which limits the achievable performance. 

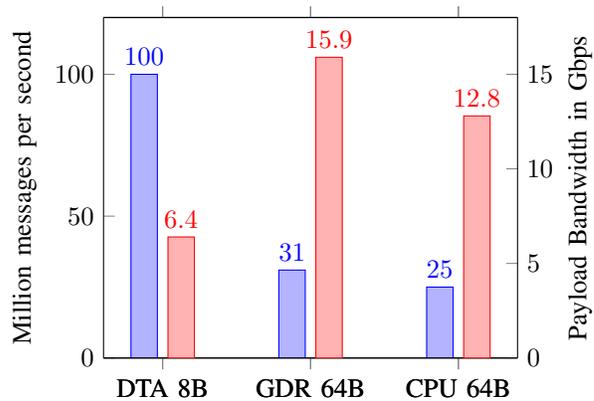
\begin{figure}
    \centering
    \begin{tikzpicture}
        \begin{axis}[ybar,
        symbolic x coords={DTA 8B, GDR 64B, CPU 64B},
        xtick=data,
        ymin=0,
        ymax=120,
        axis y line*=left,
        ylabel={Million messages per second},
        bar width=10pt,
        enlarge x limits=0.2,
        nodes near coords,
        cycle list name=color list,
        width=0.8\linewidth]

        \addplot [xshift=-7pt,color=blue,fill=blue!30] coordinates {(DTA 8B, 100) (GDR 64B, 31) (CPU 64B, 25)};
            
        \end{axis}
        \begin{axis}[ybar,
        symbolic x coords={DTA 8B, GDR 64B, CPU 64B},
        xtick=data,
        ymin=0,
        ymax=18,
        axis y line*=right,
        ylabel style={at={(axis description cs:1.1,0.5)}},
        ylabel={Payload Bandwidth in Gbps},
        bar width=10pt,
        enlarge x limits=0.2,
        nodes near coords,
        width=0.8\linewidth]

        \addplot [xshift=7pt,color=red,fill=red!30] coordinates {(DTA 8B, 6.4) (GDR 64B, 15.9) (CPU 64B, 12.8)};
            
        \end{axis}
    \end{tikzpicture}
    \vspace{-0.9em}
    \caption{Comparison of DTA 8B, to DFA with GDR 64B, and RDMA 64B with memcopy. DFA has a lower message rate, yet significantly higher payload bandwidth.}
    \vspace{-1.4em}
    \label{fig:bw-comparison}
\end{figure}

As shown in Figure~\ref{fig:bw-comparison}, the message rate reduces from 31 to 25 million feature vectors per second when the data plane sends the features to host memory using \gls{rdma} (\gls{dta}), followed by a memcopy to GPU memory. Achievable bandwidth drops accordingly. This is due to the costly memory copy operations involving the \gls{cpu}, which \gls{dfa} avoids. When comparing with the \gls{dta} 8B data transport, ~\cite{langletDirectTelemetryAccess2023} only indicates \gls{rdma} performance to host memory without GPU copy to the \gls{gpu}. As the copy performance as tested on this system, at 16 Gbps, is higher than the \gls{rdma} result achieved on system memory, when using a batched approach. This means that the whole assigned memory region is copied instead of single memory entries. When copying single 64 Byte entries, the bandwidth falls to 7 Mbps. These values were obtained by using the blocking \textit{cuMemcpyHtoD} CUDA driver function.  

%% file: discussion.tex
\section{Discussion} \label{sec:discussion}
\vspace{-0.3em}
This section provides a discussion of the practical implications and assumptions of our work. 
We do not evaluate the performance of the \gls{ml} models in terms of, such as classification accuracy. This is because we extract the same features from user traffic as Marina. Assuming the use of the same \gls{ml} models using the same 500~ms reporting interval, the ML-classification results would be similar. However, \gls{dfa} enables a 25x smaller monitoring interval compared to Marina. This may be beneficial for several AI-based network traffic-related use cases that require sub-100~ms ML-inference response times. Training new models on such smaller intervals, exploiting \gls{dfa} capabilities, is an avenue of future research.

\vspace{-0.3em}
\subsection{Performance and Deployment}
\vspace{-0.3em}
The key takeaway from our performance measurements is that \gls{dfa} supports the transfer of 31 million Marina feature vectors from the data plane directly to the \gls{gpu} memory. Assuming the ML server is co-located in the same data center as the \gls{dfa} Collectors (resulting in an \gls{rdma} latency of less than 3 ms),  feature vectors for up to 524.288 flows as in Marina, and that the additional feature derivation and pre-processing are not the bottleneck (see~\cite{seufertMarinaRealizingMLDriven2024}), \gls{dfa} supports fine-grained monitoring intervals of around 20~ms, which is a speedup of 25x compared to state-of-the-art. 
  
\vspace{1pt}
\noindent\textbf{Impact on Bottlenecks:} By sending feature vectors directly from the data plane to the ML server's \gls{gpu} using \gls{gdr}, \gls{dfa} effectively eliminates Marina's bottlenecks, which are (i) the synchronization of Marina flow features from the data plane to the control plane using costly \gls{dma} sync operations and (ii) the transmission of features to the \gls{ml} server's CPU using e.g., TCP protocol. However, while Marina implements additional feature generation by computing a set of 100 derived features as needed by the ML models on the CPU, DFA would need to implement this on CUDA kernels. As shown in~\cite{seufertMarinaRealizingMLDriven2024}, additional feature generation and ML-model execution comprise just a small portion of the total Marina processing delay, while the major part (feature sync to the control plane and transmission to the ML server) is significantly reduced when using \gls{dfa}. 

\vspace{1pt}
\noindent\textbf{Control Plane Limitations:} A bottleneck not directly addressed in our tests concerns the communication between the data plane and control plane for tracking new flows. This comes to a large proportion from the usage of Python in the control plane program, and the usage of digest messages. The original Marina control plane was implemented in the C programming language, and it clones and forwards packets that belong to flows that are not yet tracked to the control plane. It can modify 50,000 table entries (flows) per second, resulting in a flow replacement time of 6 seconds for 131,072 flows, or 20 seconds for 524,144 flows~\cite{seufertMarinaRealizingMLDriven2024}. However, \gls{dfa} for simplicity uses digest messages that are processed in Python. This results in less than 1,000 flow modifications per second. A larger share of UDP flows in relation to TCP can result in fewer flow replacement operations, as the bloom filter updates for UDP require table entry modifications, as well as contributing to additional overhead. However, a re-implementation of the control plane in C would result in similar performance as~\cite{seufertMarinaRealizingMLDriven2024}. 
    
\vspace{1pt}\noindent\textbf{Scalability of Feature Vectors:} Our measurements indicate that larger feature vectors (e.g., of 128B) do not significantly impact message rate supported by the Collector. Therefore, \gls{ai} models with larger feature sets could be supported at a similar performance, leading to potentially better ML model accuracy. Alternatively, batching multiple feature vectors into a single message could potentially double or triple the overall throughput. Such batching could be implemented in the \gls{dfa} Translator, while additional feature generation could be implemented in a \gls{snic} (e.g., in ARM cores of a  Bluefield-3). 
    
\vspace{1pt}\noindent\textbf{Hardware Deployment Considerations:} The ML-inference server that we used for the evaluation is based on the AMD EPYC architecture, which led to cross-\gls{numa} data transfers, reducing the performance. A more beneficial setup for the \gls{gpu} would be provided by a \gls{pcie} bridge, which would connect the \gls{snic} with the \gls{gpu} more directly. The difference in the 8 Byte payload test in Figures~\ref{fig:payload-size}~and~\ref{fig:bw-comparison} shows great potential for \gls{dfa} when deployed on different \gls{pcie} topologies. Especially the shallow drop-off from 8 Byte to 64 Byte payloads indicates great potential for the \gls{rdma} connection. Especially with new ML-inference server hardware architectures such as NVIDIA GraceHopper, we believe that additional speedups are possible.

    \vspace{-0.4em}
    \subsection{Security Considerations}
    \vspace{-0.3em}
        Since \gls{dfa} extracts telemetry data and directly transfers it to \gls{gpu} memory using \gls{rdma} and GPUDirect, the traditional \gls{cpu}-based security enforcement mechanisms are bypassed. This has the potential to introduce a new attack \mbox{that requires careful attention:}

        \vspace{1pt}
        \noindent\textbf{Tampering Attacks: }
            CPU-to-CPU communications commonly include protections against maliciously injected messages, such as connection-based transport, signatures, and encryption. RDMA itself typically operates with connection-based queue-pairs. However, due to hardware constraints, \gls{dta} reports are transmitted over UDP, which inherently lacks cryptographic protections, and RDMA's queue-pair security mechanisms do not safeguard incoming DTA reports before translation into RDMA.
            An adversary could therefore inject spoofed DTA reports into the system, risking corruption of ML inferences or misleading monitoring outputs.
            To preserve the hardware efficiency and low overhead of DTA, we recommend against introducing heavy cryptographic mechanisms. Instead, we propose that the DTA protocol be strictly firewalled at every network ingress point, permitting only DTA packets generated internally by the data plane. Strict enforcement of this policy would effectively eliminate the risk of malicious data injection, unless an attacker compromises and reprograms the forwarding logic within trusted switches themselves.
            Additionally, we could introduce per-reporter sequential packet identifiers. Any injected reports would then introduce duplicate packet identifiers, which would flag for abnormal telemetry behaviour to be investigated.

%% file: conclusion.tex
\vspace{-0.3em}
\section{Conclusion and Future Work} \label{sec:conclusion}
\vspace{-0.4em}
Performing ML-based traffic analysis at Terabits is challenging. In this work, we combine the approach from Marina, namely in-network feature extraction from user traffic, with \gls{dta}'s protocol extensions to ship traffic feature vectors directly from the data plane of programmable switches to \gls{gpu}s using \gls{rdma} and GPUDirect. Our work avoids costly control plane synchronization, reducing the time from collecting features to inference by a factor of 20. Bypassing the \gls{cpu} on \gls{ml} inference servers completely, we can deliver more than 30 million Marina feature vectors to the GPU \mbox{on a single network port for further analysis.}

Our work opens several avenues for future research. While feature extraction at very fine time intervals (e.g., 20~ms) is technically feasible, such granularity is often unnecessary for all \gls{ml} tasks. Future extensions could explore the adjustment of monitoring frequency based on \gls{ml} tasks. For example, for intrusion detection, low monitoring intervals would be required, while for \gls{qoe} monitoring, longer monitoring intervals would be acceptable. The monitoring interval could be even adjusted dynamically by using an additional lookup table in the Reporter devices, enabling differentiated monitoring intervals according to ML-pipelines requirements.
Another avenue for future work is batching feature vectors in the Translator or a \gls{snic}, such as NVIDIA Bluefield-3. This could alleviate the \gls{rdma} throughput bottleneck by reducing the number of transmitted messages. However, batching increases memory requirements, so an effective batching technique is desirable. 

%% file: main.bbl
\begin{thebibliography}{10}
\providecommand{\url}[1]{#1}
\csname url@samestyle\endcsname
\providecommand{\newblock}{\relax}
\providecommand{\bibinfo}[2]{#2}
\providecommand{\BIBentrySTDinterwordspacing}{\spaceskip=0pt\relax}
\providecommand{\BIBentryALTinterwordstretchfactor}{4}
\providecommand{\BIBentryALTinterwordspacing}{\spaceskip=\fontdimen2\font plus
\BIBentryALTinterwordstretchfactor\fontdimen3\font minus
  \fontdimen4\font\relax}
\providecommand{\BIBforeignlanguage}[2]{{%
\expandafter\ifx\csname l@#1\endcsname\relax
\typeout{** WARNING: IEEEtran.bst: No hyphenation pattern has been}%
\typeout{** loaded for the language `#1'. Using the pattern for}%
\typeout{** the default language instead.}%
\else
\language=\csname l@#1\endcsname
\fi
#2}}
\providecommand{\BIBdecl}{\relax}
\BIBdecl

\bibitem{claiseCiscoSystemsNetFlow2004}
B.~Claise, ``Cisco {{Systems NetFlow Services Export Version}} 9,'' Internet
  Engineering Task Force, Request for {{Comments}} RFC 3954, Oct. 2004.

\bibitem{wangSFlowResourceefficientAgile2004}
M.~Wang, B.~Li, and Z.~Li, ``{{sFlow}}: Towards resource-efficient and agile
  service federation in service overlay networks,'' in \emph{24th
  {{International Conference}} on {{Distributed Computing Systems}}, 2004.
  {{Proceedings}}.}\hskip 1em plus 0.5em minus 0.4em\relax Tokyo, Japan: IEEE,
  2004, pp. 628--635.

\bibitem{aitkenSpecificationIPFlow2013}
P.~Aitken, B.~Claise, and B.~Trammell, ``Specification of the {{IP Flow
  Information Export}} ({{IPFIX}}) {{Protocol}} for the {{Exchange}} of {{Flow
  Information}},'' Internet Engineering Task Force, Request for {{Comments}}
  RFC 7011, Sep. 2013.

\bibitem{seufertMarinaRealizingMLDriven2024}
M.~Seufert, K.~Dietz, N.~Wehner, S.~Gei{\ss}ler, J.~Sch{\"u}ler, M.~Wolz,
  A.~Hotho, P.~Casas, T.~Ho{\ss}feld, and A.~Feldmann, ``{\emph{Marina}} :
  {{Realizing ML-Driven Real-Time Network Traffic Monitoring}} at {{Terabit
  Scale}},'' \emph{IEEE Transactions on Network and Service Management},
  vol.~21, no.~3, pp. 2773--2790, Jun. 2024.

\bibitem{grayHighPerformanceNetwork2021}
N.~Gray, K.~Dietz, M.~Seufert, and T.~Hossfeld, ``High {{Performance Network
  Metadata Extraction Using P4}} for {{ML-based Intrusion Detection
  Systems}},'' in \emph{2021 {{IEEE}} 22nd {{International Conference}} on
  {{High Performance Switching}} and {{Routing}} ({{HPSR}})}.\hskip 1em plus
  0.5em minus 0.4em\relax Paris, France: IEEE, Jun. 2021, pp. 1--7.

\bibitem{salmanReviewMachineLearning2020}
O.~Salman, I.~H. Elhajj, A.~Kayssi, and A.~Chehab, ``A review on machine
  learning--based approaches for {{Internet}} traffic classification,''
  \emph{Annals of Telecommunications}, vol.~75, no. 11-12, pp. 673--710, Dec.
  2020.

\bibitem{P4}
P.~Bosshart, D.~Daly, G.~Gibb, M.~Izzard, N.~McKeown, J.~Rexford,
  C.~Schlesinger, D.~Talayco, A.~Vahdat, G.~Varghese, and D.~Walker, ``P4:
  Programming protocol-independent packet processors,'' \emph{SIGCOMM Comput.
  Commun. Rev.}, vol.~44, no.~3, pp. 87--95, Jul. 2014.

\bibitem{busse-grawitzPForestInNetworkInference2022}
C.~{Busse-Grawitz}, R.~Meier, A.~Dietm{\"u}ller, T.~B{\"u}hler, and
  L.~Vanbever, ``{{pForest}}: {{In-Network Inference}} with {{Random
  Forests}},'' Sep. 2022.

\bibitem{zhengIIsyPracticalInNetwork2022}
C.~Zheng, Z.~Xiong, T.~T. Bui, S.~Kaupmees, R.~Bensoussane, A.~Bernabeu,
  S.~Vargaftik, Y.~{Ben-Itzhak}, and N.~Zilberman, ``{{IIsy}}: {{Practical
  In-Network Classification}},'' Jun. 2022.

\bibitem{INT}
L.~Tan, W.~Su, W.~Zhang, J.~Lv, Z.~Zhang, J.~Miao, X.~Liu, and N.~Li, ``In-band
  network telemetry: A survey,'' \emph{Computer Networks}, vol. 186, p. 107763,
  2021.

\bibitem{langletDirectTelemetryAccess2023}
J.~Langlet, R.~Ben~Basat, G.~Oliaro, M.~Mitzenmacher, M.~Yu, and G.~Antichi,
  ``Direct {{Telemetry Access}},'' in \emph{Proceedings of the {{ACM SIGCOMM}}
  2023 {{Conference}}}.\hskip 1em plus 0.5em minus 0.4em\relax New York NY USA:
  ACM, Sep. 2023, pp. 832--849.

\bibitem{dimopoulosMeasuringVideoQoE2016}
G.~Dimopoulos, I.~Leontiadis, P.~{Barlet-Ros}, and K.~Papagiannaki, ``Measuring
  {{Video QoE}} from {{Encrypted Traffic}},'' in \emph{Proceedings of the 2016
  {{Internet Measurement Conference}}}, ser. {{IMC}} '16.\hskip 1em plus 0.5em
  minus 0.4em\relax New York, NY, USA: Association for Computing Machinery,
  Nov. 2016, pp. 513--526.

\bibitem{mazharRealtimeVideoQuality2018}
M.~H. Mazhar and Z.~Shafiq, ``Real-time {{Video Quality}} of {{Experience
  Monitoring}} for {{HTTPS}} and {{QUIC}},'' in \emph{{{IEEE INFOCOM}} 2018 -
  {{IEEE Conference}} on {{Computer Communications}}}, Apr. 2018, pp.
  1331--1339.

\bibitem{nvidiacorporation1OverviewGPUDirect}
{NVIDIA Corporation}, ``1. {{Overview}} --- {{GPUDirect RDMA}} 12.8
  documentation,'' https://docs.nvidia.com/cuda/gpudirect-rdma/.

\bibitem{xavierMAP4PragmaticFramework2022}
B.~M. Xavier, R.~Silva~Guimar{\~a}es, G.~Comarela, and M.~Martinello,
  ``{{MAP4}}: {{A Pragmatic Framework}} for {{In-Network Machine Learning
  Traffic Classification}},'' \emph{IEEE Transactions on Network and Service
  Management}, vol.~19, no.~4, pp. 4176--4188, Dec. 2022.

\bibitem{zhengPlanterSeedingTrees2021}
C.~Zheng and N.~Zilberman, ``Planter: Seeding trees within switches,'' in
  \emph{Proceedings of the {{SIGCOMM}} '21 {{Poster}} and {{Demo Sessions}}},
  ser. {{SIGCOMM}} '21.\hskip 1em plus 0.5em minus 0.4em\relax New York, NY,
  USA: Association for Computing Machinery, Aug. 2021, pp. 12--14.

\bibitem{lotfollahiDeepPacketNovel2018}
M.~Lotfollahi, R.~S.~H. Zade, M.~J. Siavoshani, and M.~Saberian, ``Deep
  {{Packet}}: {{A Novel Approach For Encrypted Traffic Classification Using
  Deep Learning}},'' Jul. 2018.

\bibitem{draper-gilCharacterizationEncryptedVPN2016}
G.~{Draper-Gil}, A.~H. Lashkari, M.~Mamun, and A.~Ghorbani, ``Characterization
  of {{Encrypted}} and {{VPN Traffic}} using {{Time-related Features}},''
  \emph{International Conference on Information Systems Security and Privacy},
  2016.

\bibitem{wassermannSeeWhatYou2019}
S.~Wassermann, M.~Seufert, P.~Casas, L.~Gang, and K.~Li, ``I {{See What}} you
  {{See}}: {{Real Time Prediction}} of {{Video Quality}} from {{Encrypted
  Streaming Traffic}},'' in \emph{Proceedings of the 4th {{Internet-QoE
  Workshop}} on {{QoE-based Analysis}} and {{Management}} of {{Data
  Communication Networks}}}.\hskip 1em plus 0.5em minus 0.4em\relax Los Cabos
  Mexico: ACM, Oct. 2019, pp. 1--6.

\bibitem{10.5555/2930611.2930632}
Y.~Li, R.~Miao, C.~Kim, and M.~Yu, ``{{FlowRadar}}: A better {{NetFlow}} for
  data centers,'' in \emph{Proceedings of the 13th Usenix Conference on
  Networked Systems Design and Implementation}, ser. {{NSDI}}'16.\hskip 1em
  plus 0.5em minus 0.4em\relax Santa Clara, CA and USA: USENIX Association,
  2016, pp. 311--324.

\bibitem{10.1145/3190508.3190558}
J.~Sonchack, A.~J. Aviv, E.~Keller, and J.~M. Smith, ``Turboflow: Information
  rich flow record generation on commodity switches,'' in \emph{Proceedings of
  the Thirteenth {{EuroSys}} Conference}, ser. {{EuroSys}} '18.\hskip 1em plus
  0.5em minus 0.4em\relax Porto, Portugal and New York, NY, USA: Association
  for Computing Machinery, 2018.

\bibitem{9495221}
Z.~Zhao, X.~Shi, Z.~Wang, Q.~Li, H.~Zhang, and X.~Yin, ``Efficient and accurate
  flow record collection with {{HashFlow}},'' \emph{IEEE Transactions on
  Parallel and Distributed Systems}, vol.~33, no.~5, pp. 1069--1083, 2022.

\bibitem{265057}
Q.~Huang, S.~Sheng, X.~Chen, Y.~Bao, R.~Zhang, Y.~Xu, and G.~Zhang, ``Toward
  {{Nearly-Zero-Error}} sketching via compressive sensing,'' in \emph{18th
  {{USENIX}} Symposium on Networked Systems Design and Implementation ({{NSDI}}
  21)}.\hskip 1em plus 0.5em minus 0.4em\relax USENIX Association, Apr. 2021,
  pp. 1027--1044.

\bibitem{10.1145/3230543.3230544}
T.~Yang, J.~Jiang, P.~Liu, Q.~Huang, J.~Gong, Y.~Zhou, R.~Miao, X.~Li, and
  S.~Uhlig, ``Elastic sketch: Adaptive and fast network-wide measurements,'' in
  \emph{Proc. of the {{ACM SIGCOMM}} 2018}.\hskip 1em plus 0.5em minus
  0.4em\relax Budapest, Hungary and New York, NY, USA: Association for
  Computing Machinery, 2018, pp. 561--575.

\bibitem{10.14778/3467861.3467868}
S.~Sheng, Q.~Huang, S.~Wang, and Y.~Bao, ``{{PR-sketch}}: Monitoring per-key
  aggregation of streaming data with nearly full accuracy,'' \emph{Proc. VLDB
  Endow.}, vol.~14, no.~10, pp. 1783--1796, Jun. 2021.

\bibitem{10.1145/3626775}
A.~Monterubbiano, J.~Langlet, S.~Walzer, G.~Antichi, P.~Reviriego, and
  S.~Pontarelli, ``Lightweight acquisition and ranging of flows in the data
  plane,'' \emph{Proc. ACM Meas. Anal. Comput. Syst.}, vol.~7, no.~3, Dec.
  2023.

\bibitem{9064137}
J.~Vestin, A.~Kassler, D.~Bhamare, K.-J. Grinnemo, J.-O. Andersson, and
  G.~Pongracz, ``Programmable event detection for in-band network telemetry,''
  in \emph{2019 {{IEEE}} 8th International Conference on Cloud Networking
  ({{CloudNet}})}, 2019, pp. 1--6.

\bibitem{Kim2015InbandNT}
C.~Kim, A.~Sivaraman, N.~P.~K. Katta, A.~Bas, A.~A. Dixit, and L.~J. Wobker,
  ``In-band network telemetry via programmable dataplanes,'' \emph{Proc. ACM
  SIGCOMM 2015}, 2015.

\bibitem{265015}
Y.~Zhao, K.~Yang, Z.~Liu, T.~Yang, L.~Chen, S.~Liu, N.~Zheng, R.~Wang, H.~Wu,
  Y.~Wang, and N.~Zhang, ``{{LightGuardian}}: A {{Full-Visibility}},
  lightweight, in-band telemetry system using sketchlets,'' in \emph{18th
  {{USENIX}} Symposium on Networked Systems Design and Implementation ({{NSDI}}
  21)}.\hskip 1em plus 0.5em minus 0.4em\relax USENIX Association, Apr. 2021,
  pp. 991--1010.

\bibitem{10.1145/3387514.3405877}
Q.~Huang, H.~Sun, P.~P.~C. Lee, W.~Bai, F.~Zhu, and Y.~Bao, ``{{OmniMon}}:
  {{Re-architecting}} network telemetry with resource efficiency and full
  accuracy,'' in \emph{Proc. {{ACM SIGCOMM}} 2020}.\hskip 1em plus 0.5em minus
  0.4em\relax Virtual Event, USA and New York, NY, USA: Association for
  Computing Machinery, 2020, pp. 404--421.

\bibitem{Cisco}
{Cisco}, ``Explore model-driven telemetry,'' 2019, accessed: 2025-04-26.

\bibitem{Juniper}
{Juniper}, ``Overview of the junos telemetry interface,'' 2021, accessed:
  2025-04-26.

\bibitem{Telemetry}
{The P4.org Applications Working Group}, ``Telemetry report format
  specification,'' 2020.

\bibitem{9447791}
E.~F. Kfoury, J.~Crichigno, and E.~{Bou-Harb}, ``An exhaustive survey on {{P4}}
  programmable data plane switches: {{Taxonomy}}, applications, challenges, and
  future trends,'' \emph{IEEE access : practical innovations, open solutions},
  vol.~9, pp. 87\,094--87\,155, 2021.

\bibitem{10.1145/3387514.3406214}
Y.~Zhou, C.~Sun, H.~H. Liu, R.~Miao, S.~Bai, B.~Li, Z.~Zheng, L.~Zhu, Z.~Shen,
  Y.~Xi, P.~Zhang, D.~Cai, M.~Zhang, and M.~Xu, ``Flow event telemetry on
  programmable data plane,'' in \emph{Proc. {{ACM SIGCOMM}} 2020}.\hskip 1em
  plus 0.5em minus 0.4em\relax Virtual Event, USA and New York, NY, USA:
  Association for Computing Machinery, 2020, pp. 76--89.

\bibitem{10.1145/3098822.3098829}
S.~Narayana, A.~Sivaraman, V.~Nathan, P.~Goyal, V.~Arun, M.~Alizadeh,
  V.~Jeyakumar, and C.~Kim, ``Language-directed hardware design for network
  performance monitoring,'' in \emph{Proc. {{ACM SIGCOMM}} 2017}.\hskip 1em
  plus 0.5em minus 0.4em\relax Los Angeles, CA, USA and New York, NY, USA:
  Association for Computing Machinery, 2017, pp. 85--98.

\bibitem{287396}
\BIBentryALTinterwordspacing
M.~Shen, K.~Ji, Z.~Gao, Q.~Li, L.~Zhu, and K.~Xu, ``Subverting website
  fingerprinting defenses with robust traffic representation,'' in \emph{32nd
  USENIX Security Symposium (USENIX Security 23)}.\hskip 1em plus 0.5em minus
  0.4em\relax Anaheim, CA: USENIX Association, Aug. 2023, pp. 607--624.
  [Online]. Available:
  \url{https://www.usenix.org/conference/usenixsecurity23/presentation/shen-meng}
\BIBentrySTDinterwordspacing

\bibitem{edgecorenetworkscorporationDCS801Wedge100BF32QSDS2023}
{Edgecore Networks Corporation}, ``{{DCS801}} ({{Wedge100BF-32QS}}) {{DS
  R05}},'' 2023.

\bibitem{TRex}
Cisco, ``{T-Rex},'' https://trex-tgn.cisco.com/, 2022, accessed: 2025-04-26.

\bibitem{IntelP4Insight}
{Intel}, ``{Intel{\textregistered} P4 Insight Tool},''
  \url{https://www.intel.com/content/www/de/de/products/details/network-io/intelligent-fabric-processors/p4-insight.html},
  accessed: 2025-04-26.

\end{thebibliography}
